\documentclass{webofc}
\usepackage[varg]{txfonts}   
\begin{document}
\title{Charm and multi-charm baryon measurements via
strangeness tracking with the upgraded ALICE detector}
%
%

\author{\firstname{David} \lastname{Dobrigkeit Chinellato for the ALICE Collaboration}\inst{1,2}\fnsep\thanks{\email{daviddc@ifi.unicamp.br}}
}

\institute{UNICAMP
\and
           CERN
          }

\abstract{%

We  present a new method for detection of multiply charmed
baryons via their decays into strange baryons, using `strangeness 
tracking'. This method makes use of the state-of-the-art upgraded silicon detectors in ALICE during Runs 3, 4 and beyond will enable the novel possibility of tracking strange hadrons directly before they decay, leading to a very significant improvement in impact-parameter resolution. In this work, we will discuss how this new technique will be crucial to distinguish secondary strange baryons originating from 
charm decays from primary strange baryons. This is a particularly
interesting possibility for the $\Omega^{-}$ baryon coming
from $\Omega_{\rm{c}}^{0}\rightarrow\Omega^{-}\pi^{+}$ decays, since there is no other relevant feeddown source for $\Omega^{-}$. This, in turn, means that the main $\Omega^{-}$ background for the $\Omega_{\rm{c}}$ measurement will point most accurately to the primary vertex, unlike pions or protons from other charm baryon decays. 

We will illustrate the achievable performance of strangeness tracking for the Run 3 configuration of ALICE with the upgraded Inner Tracking System, which is fully instrumented with silicon pixel detectors. Moreover, we will discuss the potential of this technique in a future experiment with an extensive silicon tracking detector with a first layer very close to the interaction point.
Finally, we will also cover other potential major applications 
of strangeness tracking, including measurements of hypernuclei such as the $^{3}_{\Lambda}\rm{H}$.

}
\maketitle
\section{Introduction}
\label{intro}

A fundamental ingredient of the ALICE physics programme for the new decade is a comprehensive study of 
charm and multi-charm baryon production. Because charm is exclusively produced in initial hard scatterings, 
such measurements may provide unique insight into the QGP medium as well as hadronization from proton-proton
to lead-lead collisions. However, the reconstruction of charmed hadrons poses a significant challenge. In particular, charm particles decay into strange particles, in which case topological reconstruction can 
serve as 
a powerful particle identification method. However, due to the large extrapolation distances, weak decay reconstruction is usually also associated with less than optimal pointing resolution to the primary vertex. 

\begin{figure}[h]
\centering
\includegraphics[width=12cm]{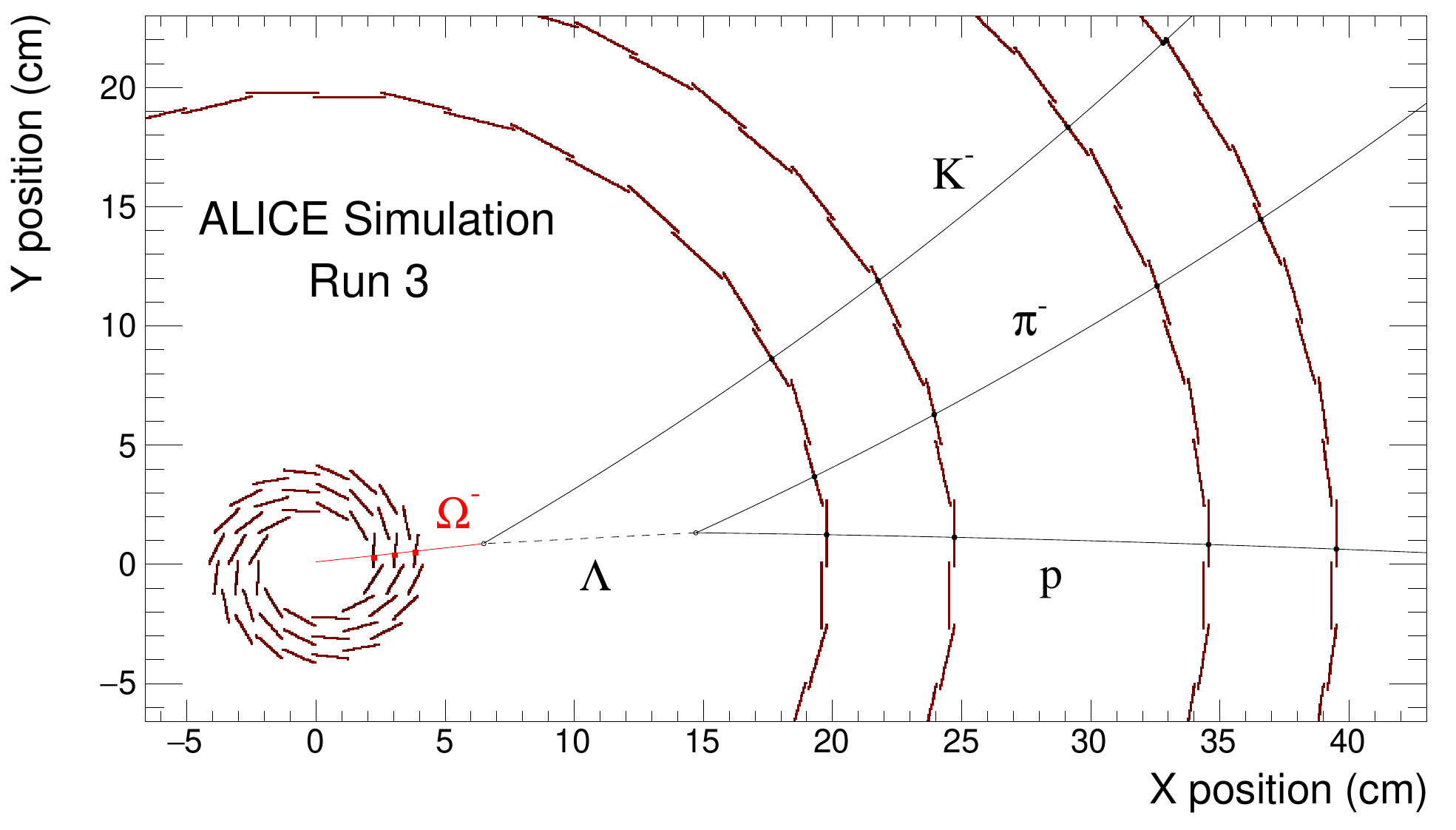}
\caption{Event display of an $\Omega^{-}\rightarrow K^{-}+\Lambda$ decay in the 
upgraded ALICE detector in Run 3. The $\Omega^{-}$ hits in the inner layers 
of the ALICE ITS2 are highlighted in red and are used in the reconstruction of 
the weak decay when the strangeness tracking technique is employed. }
\label{fig-1}       
\end{figure}

We propose to overcome this
difficulty using a novel reconstruction technique called `strangeness tracking', which is made possible by the 
next-generation detectors of the experiment. In this method, we use high-resolution silicon detectors very close 
to the primary vertex to measure the hits of a weakly decaying hadron prior to its decay and combine them with
the information from the decay daughters to significantly improve pointing resolution. 
In this context, the upgraded ALICE inner tracking system, the ITS2 \cite{its2tdr}, is ideal for a first implementation 
of such a method. This is because the ITS2 has an innermost silicon 
pixel layer at a radius of 22~mm, as opposed to 39~mm for the previous silicon pixel detectors. 
As a consequence, the two multi-strange baryons $\Xi^{-}$ and $\Omega^{-}$, reconstructed in the channels $\Xi^{-}\rightarrow \pi^{-}+\rm{p}$ and
$\Omega^{-}\rightarrow \rm{K}^{-}+\rm{p}$ channels and having a proper lifetime of 4.9 and 2.5~cm, respectively, 
are significantly more likely to leave hits in the innermost layers of the ITS2. An example of a reconstructed $\Omega^{-}$ decay obtained with a simulation of the full geometry of the upgraded ALICE detector and using GEANT3  propagation is shown in Fig.~\ref{fig-1}. 

\section{Feasibility of strangeness tracking}
\label{Feasibility}

Algorithmically, the task of performing strangeness tracking involves a precise enough reconstruction of the 
decay daughters such that the back propagation of a $\Xi^{-}$ or $\Omega^{-}$ candidate is 
sufficiently constraining for the association of the correct inner layer hits to this trajectory. This establishes a crucial 
performance constraint: in order to ensure a low fake hit association rate, the hit density in a given layer, 
$\rho_{hits}$, should obey the relation 
$1/\rho_{hits} \gg \delta_{\rm{search}}$, where $\delta_{\rm{search}}$ is the search window that is used 
when propagating the cascade trajectory inward to that layer. The parameter $\delta_{\rm{search}}$ has been calculated in 
simulations and is below 0.05~$\rm{mm}^{2}$ for typical $\Omega^{-}$ and $\Xi^{-}$ inwards propagation in the 
innermost layer, while hit densities are not expected to exceed 50~$\rm{cm}^{-2}$ even in central Pb--Pb collisions. Because 
$1/\rho_{hits} = 2$~$\rm{mm}^{2} \gg 0.05$~$\rm{mm}^{2}$, no major difficulties are expected when implementing
strangeness tracking for Run 3 data. In this study, Monte Carlo 
association of the cascade hits will be employed to illustrate the expected gains of the strangeness tracking method 
in an idealised scenario. 

\section{Reconstruction improvements}
\label{Improvement}

\begin{figure*}
\centering
\includegraphics[width=0.475\textwidth]{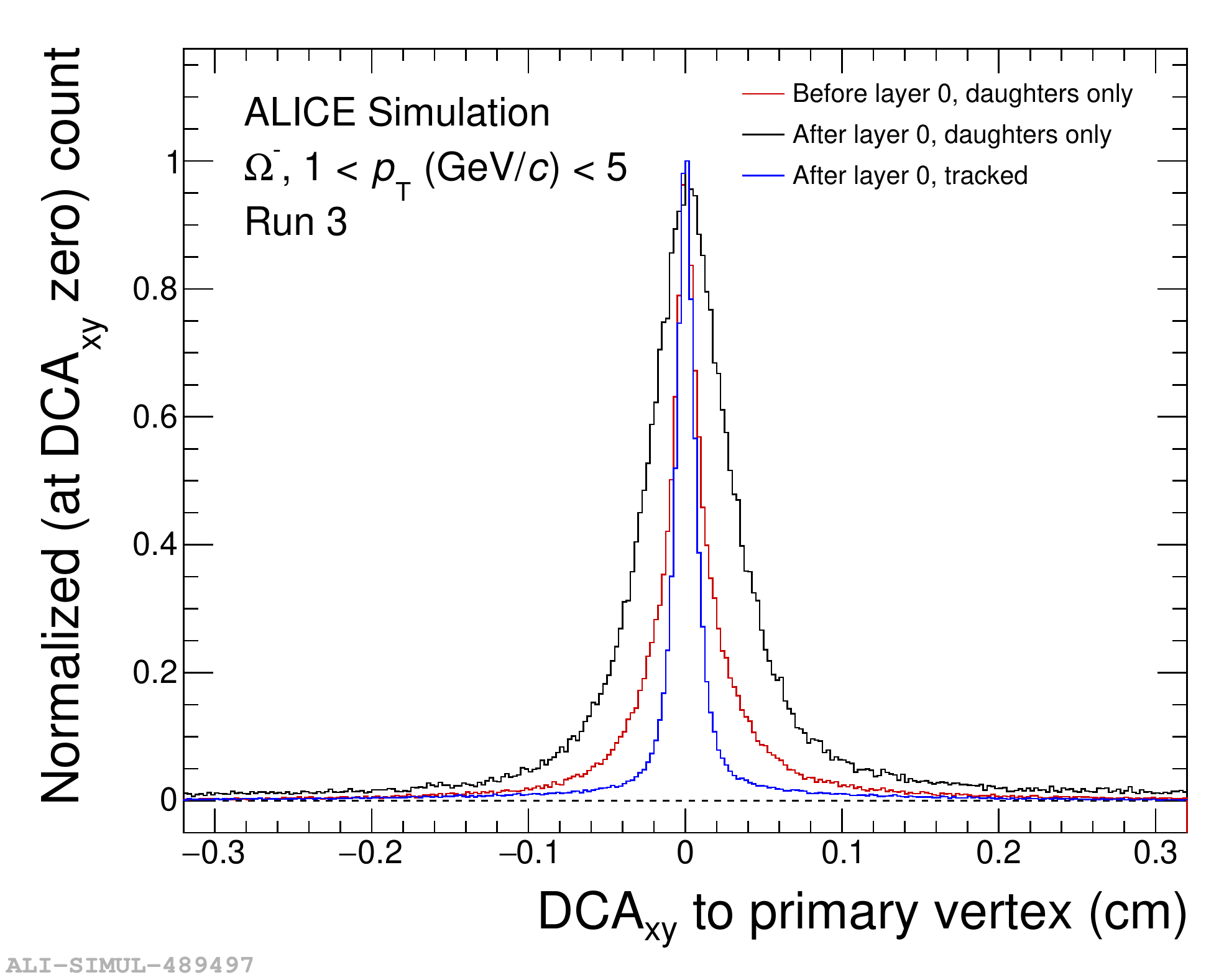}
\includegraphics[width=0.475\textwidth]{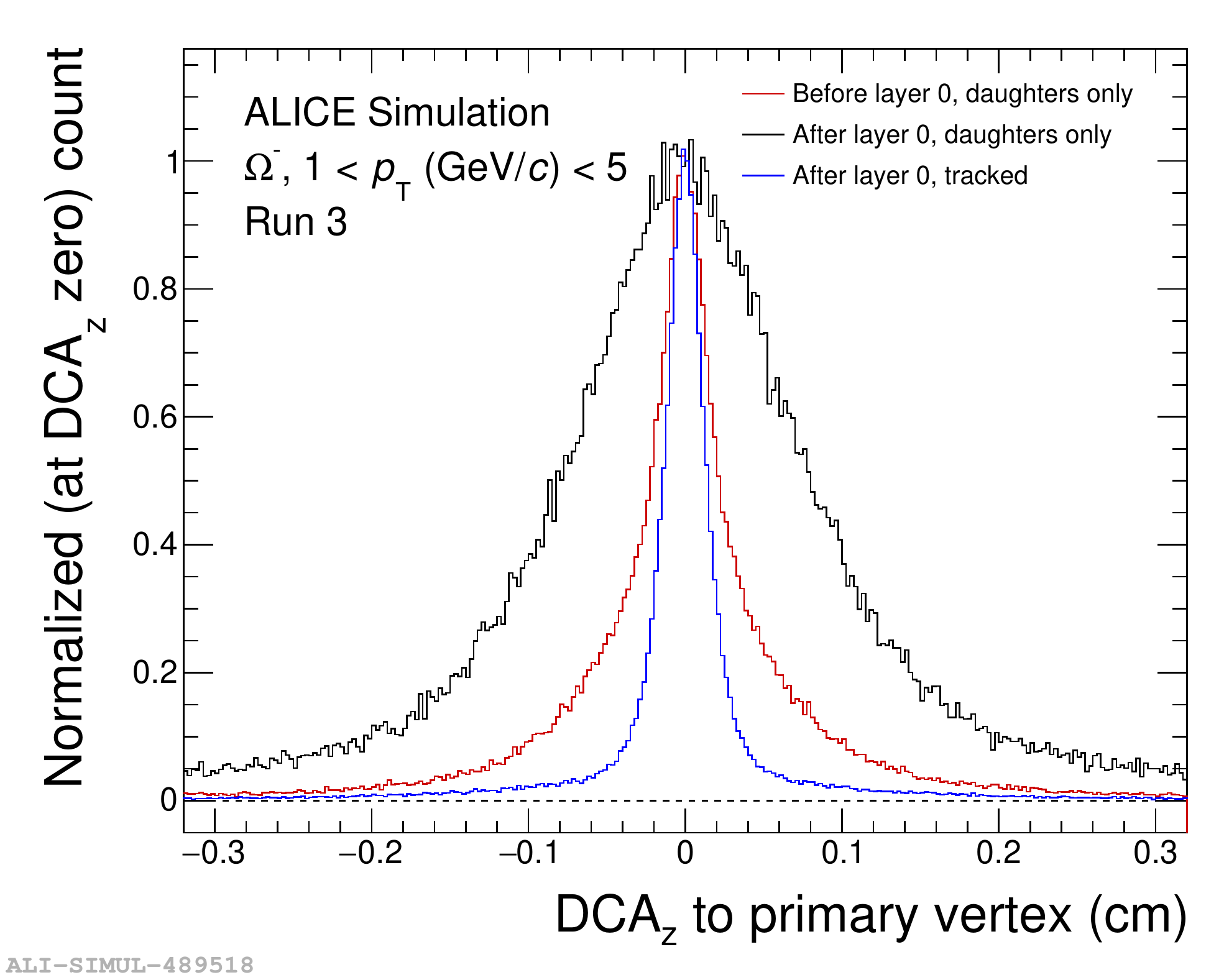}
\caption{Distance of closest approach to the primary vertex for $\Omega^{-}$ baryons reconstructed
in proton-proton collisions at 13 TeV with the upgraded ALICE detector in Run 3 in three different configurations, as obtained with a full detector simulation.}
\label{fig-2}       
\end{figure*}

The direct tracking of weakly decaying hadrons is expected to have a large impact in the pointing resolution, 
which can, for instance, be readily quantified by investigating the effect on the distribution of distance of closest approach (DCA)
of $\Omega^{-}$ baryons to the primary vertex. This is shown in Fig.~\ref{fig-2} for the $\rm{DCA}_{\rm xy}$ and the $\rm{DCA}_{\rm z}$, which are measured in the transverse plane and in the longitudinal direction, respectively. Three distributions are shown: 
(a) one in which the $\Omega^{-}$ decayed prior to the first layer and has therefore only
been reconstructed using daughter information, (b) a second curve in which the $\Omega^{-}$ decayed after the first 
layer but no direct measurement was performed, illustrating the significant worsening of the DCA resolution and (c) 
a third curve in which $\Omega^{-}$ baryons decay after the first layer and direct detection information is added
to the daughter information. The resulting improvement in DCA is of a factor of approximately 4 in both transverse and longitudinal directions, showcasing the potential of the method. 

\begin{figure*}
\centering
\includegraphics[width=0.55\textwidth]{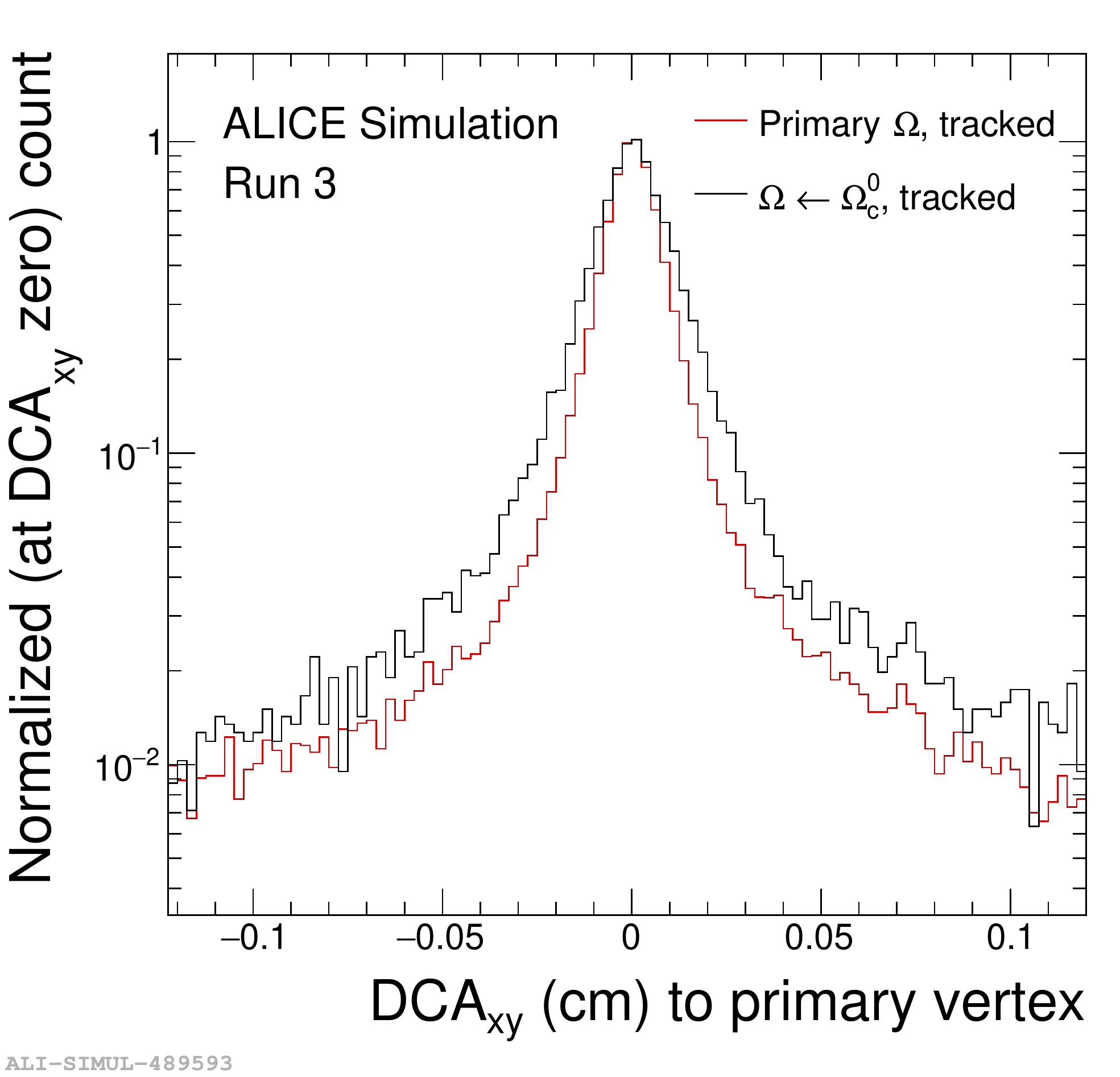}
\caption{Distance of closest approach to the primary vertex for primary $\Omega^{-}$ baryons and $\Omega^{-}$ from 
$\Omega_{c}^{0}\rightarrow \Omega^{-}+\pi^{+}$ decays as reconstructed
in proton-proton collisions at 13 TeV with the upgraded ALICE detector in Run 3.}
\label{fig-3}       
\end{figure*}

When used for the detection of $\Omega^{-}$ from $\Omega_{\rm{c}}^{0}$ decays, strangeness tracking is 
crucial to isolate differences in the pointing to the primary vertex between these two sources, which 
will be due to the lifetime of the $\Omega_{\rm{c}}^{0}$, which has $c\tau$~=~80~$\mu \rm{m}/c$, as illustrated in Fig.~\ref{fig-3}.  
While for the ITS2 
the differences between primary and secondary $\Omega^{-}$ baryons are still not too evident, this is 
expected to improve significantly with further upgrades to the ALICE apparatus, which will enable a better 
DCA to primary vertex resolution for primary $\Omega^{-}$; such upgrades will be discussed
in the last section of this document. 

Similar tests were also performed for the $\Xi^{-}$ baryon. The DCA resolution increase is comparable 
to the $\Omega^{-}$ one and strangeness tracking also outperforms the situation in which the $\Xi^{-}$  decays 
before the first silicon layer and therefore all daughters are tracked as best as possible. Moreover, early tests 
with hypertriton reconstruction also indicate a significant improvement in pointing performance that may also serve 
to greatly reduce the background in hypertriton analyses, as DCA selections help in removing background from 
secondary deuterons. 

\section{Outlook}
\label{Outlook}

As studied in this work, strangeness tracking promises to dramatically improve the pointing resolution of 
weakly decaying particles such as the $\Xi^{-}$, $\Omega^{-}$ and even hypernuclei. However, 
further upgrades will enable even better performance. Notably, the upgrade of the ITS scheduled for the third long
shutdown of the LHC will bring the innermost layers even closer to the interaction vertex with the ITS3 \cite{its3loi} 
and reduce material budget significantly, 
further improving both the performance and the expected fiducial for strangeness tracking. Internal testing
of strangeness tracking in that setup has already begun and shows significant gains. 

Further in the future, a next-generation, all-silicon apparatus has been proposed as a replacement for the TPC-based
ALICE experiment \cite{alice3}. For this new setup, a design is being considered that has a retractable silicon layer that is positioned only 5~mm 
away from the beam axis during datataking and would serve as the ultimate strangeness tracking tool. A cornerstone
of the physics programme of this setup would be the detection of the multi-charm baryons $\Xi_{\rm{cc}}^{++}$, $\Omega_{\rm{cc}}^{+}$ and $\Omega_{\rm{ccc}}^{++}$, all of which have decay channels into multi-strange baryons. Also in this case, strangeness
tracking studies are ongoing and the ability to separate signal from background will be reported very soon. 

%
%
%

\end{document}